\def\aa{{A\&A}}
\def\aap{{A\&A}}
\def\aaps{{A\&AS}}
\def\aas{{A\&AS}}
\def\aj{{AJ}}
\def\annrev{{ARA\&A}}
\def\apj{{ApJ}}
\def\apjs{{ApJS}}
\def\baas{{BAAS}}
\def\mnras{{MNRAS}}
\def\nat{{Nature}}
\def\pasp{{PASP}}

\documentclass[cup5b]{caps}
\usepackage{graphicx}
\usepackage{amssymb}
\usepackage{ociwsymp4e}  
\HeadText{Kraft \& Ivans}

\begin{document}

\pagenumbering{arabic}

%
\author[]{R.\ P.\ KRAFT$^{1}$ and I.\ I.\ IVANS$^{2}$
\\
(1) UCO/Lick, University of California, Santa Cruz, CA, USA\\
(2) California Institute of Technology, Pasadena, CA, USA}

\chapter{A New Globular Cluster Metallicity Scale Based on the Abundance of Fe~\sc{ii}}

\begin{abstract}
When using Fe as a surrogate for ``metallicity'', the metallicity is
best represented by the dominant species of Fe.  Accordingly, we have 
derived a new globular cluster metallicity scale based on the equivalent 
widths of Fe~{\sc ii} lines measured from high resolution spectra of 
giant stars.  The scale is primarily based on the results of analyses by 
the Lick-Texas group of 149 stars in 11 clusters, supplemented by other
high resolution studies in five additional clusters.  We also derive 
{\it ab initio} the true distance moduli for M3, M5, M13, M15, and M92 
as a means of setting stellar surface gravities.  We find that 
[Fe/H]$_{\rm II}$ is correlated linearly with $W'$, the reduced strength 
of the near-infrared Ca~{\rm II} triplet defined by Rutledge et al 
(1997), although the correlation coefficients depend on the stellar 
atmosphere model employed.  In addition to the 66 globular cluster 
metallicity estimates presented in a recent PASP review, we present here 
an additional 39 globular cluster metallicity estimates based on 
transformations from Q39, the photometric index defined by Zinn (1980).
\end{abstract}

\section{Background}

Because globular cluster stars are faint, they are difficult to study
at high spectral resolution.  Thus, metallicity-sensitive indices have
been devised from photometric and low- to medium-resolution data in
order to study large samples of clusters.  Zinn \& West (1984; ``ZW84'') 
provided the first truly extensive data set based on 
metallicity-sensitive photometric indices of integrated cluster light.
Most recently, Rutledge et al (1997a,b), employed infra-red (``IR'')
Ca~{\sc ii} triplet features in low-resolution spectra of individual
cluster giant stars.  Regardless of the method employed, both indices
require calibration to high-resolution abundance analyses of [Fe/H].
Metallicity estimates for other clusters are then derived by 
interpolating the correlation of the observed values of the indices
and [Fe/H] determinations.  The resulting metallicity scale therefore 
depends upon the reliability of the [Fe/H] determinations of nearby
key clusters.

At the present time, two competing calibrations based on high
resolution spectroscopic analyses exist.  ZW84 employed the
high resolution studies of Cohen (1978, 1979, 1980, 1981) and
supplemented the results by Butler's (1975) measurements of 
$\Delta$$S$ determinations for RR Lyrae stars.  Later, in a
study by Carretta \& Gratton (1997; ``CG97''), CCD-based 
equivalent width (``EW'') measurements were adjusted to a common
system in order to define a new metallicity scale.  CG97 employed a 
common set of Fe~{\sc i} {\it gf}-values, Kurucz (1992, 1993) model 
atmospheres with overshooting, and a uniform colour-effective 
temperature (``T$_{\rm eff}$'') relationship (Gratton et al 1996) 
for 160 stars in 24 globular clusters.  The difference in the 
metallicity scales, in the sense of $\Delta$[Fe/H] = {\it CG97} -- 
{\it ZW84}, is --0.1~dex for metal-rich clusters and $\sim$ 
+0.2~dex for metal-poor clusters.

\section{Developments Since 1997}

In the five years since the CG97 scale was established, there have
been new developments that affect the abundance analysis of 
metal-poor giants, and thus the metallicity scale of globular
clusters.  First, new scales of colour versus T$_{\rm eff}$ have been 
devised, one by Alonso et al (1999) based on the IR Flux Method 
(Blackwell et al 1990), and another by Houdashelt et al (2000) who 
modelled broad-band colours as a function of T$_{\rm eff}$ and 
stellar surface gravity (``log~{\it g}'').  Second, consideration of 
the effects of non-local thermodynamic equalibrium (``non-LTE'') 
suggests that the assumption of LTE ionization equilibrium between 
Fe~{\sc i} and Fe~{\sc ii} may be invalid; iron may be overionized 
in the atmospheres of metal-poor giant stars (see e.g.\ Th\'evenin 
\& Idiart 1999; ``TI99'').  Finally, new cluster distance moduli
based on {\it Hipparcos} (ESA 1997) parallaxes for nearby subdwarfs 
has led to slightly increased globular cluster distances, which has
slightly decreased the estimates of log~{\it g} in globular cluster 
stars.

\section{Why Fe~{\sc ii}?}

By ``metallicity'', we make the usual assumption that it is 
equivalent to [Fe/H] as derived from high resolution analyses.  In the 
past, [Fe/H] has normally been based on the abundance derived from 
Fe~{\sc i} lines, or on the mean of the abundances derived from both 
Fe~{\sc i} and Fe~{\sc ii} lines.  However, in a detailed study of LTE 
models of metal-poor stellar atmospheres, TI99 concluded that abundance 
analyses of Fe~{\sc i} underestimate the abundance of iron.  At any 
given optical depth, the populations of the atomic levels of Fe~{\sc i} 
are affected by the outward leakage of higher energy photons from inner 
layers; the local kinetic temperature is not in equilibrium with the 
radiation field.  The result is that iron is overionized.  But, in such
atmospheres, Fe~{\sc ii} is by far the dominant species and is
essentially unaffected by departures from LTE.  So, the effect of this 
lack of equilibrium, while serious for Fe~{\sc i}, is negligible for 
Fe~{\sc ii}.  Thus, metallicities for globular clusters can safely be 
based on LTE analysis of Fe~II.  This is supported by 3-dimensional 
atmosphere modelling by Asplund \& Garcia Perez (2001) and Nissen et al 
(2002).

\section{Basis of Re-Analysis}

Although the analyses vary widely in cluster sample size, high 
resolution abundance analyses for giant stars in over 30 globular 
clusters can be found in the literature.  Many giants are in common to 
multiple investigations.  These stars provide information from which 
T$_{\rm eff}$ and log~{\it g} transformations can be estimated.  We 
focussed our attention on those clusters in which a reasonable number 
of Fe~{\sc ii} lines ($N >$4) were measured in each giant star analyzed 
and to those clusters in which at least three giants were studied.  The 
clusters studied by the Lick-Texas group (see Kraft \& Ivans 2003 for 
all of the references employed; ``KI-03'') usually proved to include 
results for more stars in a given cluster than those included by other 
groups.  In addition, the Fe linelists and {\it gf}-values chosen 
varied considerably from group to group.  Thus, we focussed our 
attention on 16 clusters, uniformly analyzed for Fe~{\sc ii} abundances 
mostly by the Lick-Texas group, ranging in metallicity --2.5 $\le$ 
[Fe/H]$_{\rm II}$ $\le$ --0.7.  The present results are based on more 
than 150 stars.

In order to put the clusters on the same metallicity scale, we needed 
to make adjustments to the input parameters.  The following 
considerations were incorporated:  
(i) the adopted solar Fe abundance; 
(ii) the derivation of {\it gf}-values for Fe~{\sc i} and Fe~{\sc ii}; 
(iii) estimates of interstellar reddening; 
(iv) the assignment of T$_{\rm eff}$; 
(v) estimates of cluster moduli and stellar masses; and
(vi) the determination of log~{\it g}.  The Lick-Texas group studies,
which made up the bulk of our sample, are based on analyses employing 
MARCS models (Gustafsson et al 1975).  Thus, we also derived 
transformations to Kurucz models, both with and without overshooting.
We adopted the colour-T$_{\rm eff}$ relationship of Alonso et al but 
also derived the offsets that need to be applied if one employs the 
Houdashelt et al calibration instead.  Details of the strategy we 
devised are found in KI-03 (their \S3 and Table~5).

In column (3) of Table~\ref{table1}, we display the true distance 
moduli we derived for five key clusters on the basis of 
{\it Hipparcos}-based distances to low-metallicity subdwarfs.  
Column (2) shows the metallicity we derived.  Details regarding the 
method of the analysis, including the adoption of T$_{\rm eff}$, 
log~{\it g}, {\it gf}-values, the solar model, and 
log~$\epsilon$(Fe)$_{\odot}$, are presented in KI-03.

\begin{table}
\caption{True Distance Moduli of Five Key Clusters}
\begin{tabular}{lccl}
\hline \hline
     {Cluster} & {[Fe/H]$_{\rm II}$}    & {$(m$$-$$M)_{0}$}    & {$(m$$-$$M)_{0}$ from Previous Studies}\\
        (1)    &       (2)              &        (3)           &       (4) \\
     \hline
M15            & --2.42            & 15.25             & 15.26 (1) 
                                                         15.25 (2a) 
                                                         15.31 (2b)  \\
M92            & --2.38            & 14.75             & 14.49 (1)
                                                         14.93 (2)
                                                         14.71 (3) \\
M13	       & --1.60            & 14.42             & 14.08 (1c)
                                                         14.26 (1d)
                                                         14.48 (2)
                                                         14.40 (3) \\
M3	       & --1.50		   & 15.02	       & 14.79 (1) \\
M5	       & --1.26		   & 14.42	       & 14.03 (1e)
						         14.40 (1f)
                                                         14.45 (2)
                                                         14.59 (3) \\
\hline \hline
\end{tabular}

\hspace{8pt} References:\\
(1) Lick-Texas group (all relevant citations can be found in in KI-03)\\
(2) Reid 1997\\
(3) Carretta et al 2000\\
\hspace{8pt} Additional Notes:\\
(a) Reid $E(B$$-$$V)$ = 0.07\\
(b) Reid $E(B$$-$$V)$ = 0.11\\
(c) Kraft et al 1997\\
(d) Pilachowski et al 1996\\
(e) Sneden et al 1992\\
(f) Ivans et al 2001
\label{table1}
\end{table}

\section{Revised Cluster Abundances Based on Fe~{\sc ii}}

The clusters in our sample were divided into three groups.  Group 1
clusters are those with T$_{\rm eff}$ and log~{\it g} based on colours
and absolute magnitudes.  They all possess $E(B$$-$$V)$ $\le$ 0.10
and Lick-Texas group analyses of at least six giants per cluster.
Group 2 clusters are those with larger, less certain reddenings.
Group 3 clusters include those from non-Lick-Texas studies with the 
results transformed to those on the Lick-Texas group system.  
Table~\ref{table2} presents the reddening and distance moduli values 
we adopted in our study, as well as the revised cluster abundances 
based on Fe~{\sc ii}.  For each cluster, three metallicity estimate 
values are provided.  The estimates in column (4) corresponds to 
those based on the Lick-Texas group analysis which employed MARCS 
stellar atmosphere models.  Columns (6) and (7) correspond to those 
derived in the Lick-Texas system and transformed to systems based on 
Kurucz stellar atmosphere models, with (``K-on'') and without 
(``K-off'') overshooting (we refer the reader to KI-03, their \S5, 
for a detailed discussion of the transformation).  Owing to the 
method we employed, the $\sigma$ reported in column (5) is the same
for columns (4), (6), and (7).  The Fe~{\sc ii} metallicities in 
Table~\ref{table2} were derived employing the Alonso et al 
colour-T$_{\rm eff}$ calibrations.  The difference between the 
Alonso et al and Houdashelt et al T$_{\rm eff}$ scales induces a 
difference of $\sim$ --0.04~dex in [Fe/H]$_{\rm II}$ (we refer the 
reader to KI-03, their \S6 and appendix, for details).

\begin{table}
\caption{Revised Globular Cluster Abundances Based on Individual Stellar Abundances of Fe~{\sc ii}, Segregated By Model for 16 Key Clusters}
\begin{tabular}{lcccccc}
\hline \hline
     {Cluster} & {$E(B$$-$$V)$} & $(m$$-$$M)_{0}$ & [Fe/H]$_{MARCS}$ & {$\sigma$} & [Fe/H]$_{K-on}$ & [Fe/H]$_{K-off}$ \\
        (1)    &      (2)       &      (3)        &         (4)      &    (5)     &    (6)          &(7) \\
     \hline
 	 & 	 & 	  & 	Group 1 & & & \\
 M5      & 0.03  & 14.42 & --1.26 & 0.06 & --1.19 & --1.19 \\
N362     & 0.05  & 14.70 & --1.34 & 0.07 & --1.27 & --1.27 \\
N288     & 0.03  & 14.66 & --1.41 & 0.04 & --1.33 & --1.35 \\
 M3      & 0.01  & 15.02 & --1.50 & 0.03 & --1.42 & --1.43 \\
 M13     & 0.02  & 14.42 & --1.60 & 0.08 & --1.52 & --1.53 \\
 M92     & 0.02  & 14.75 & --2.38 & 0.07 & --2.32 & --2.38 \\
 M15     & 0.10  & 15.25 & --2.42 & 0.07 & --2.36 & --2.42 \\
 	 & 	 & 	  & 	Group 2 & & & \\
 M71     & 0.32  & 12.83 & --0.81 & 0.07 & --0.74 & --0.74 \\
 M4(a)   & 0.33  & 11.61 & --1.15 & 0.08 & --1.08 & --1.08 \\
 M10     & 0.24  & 13.41 & --1.51 & 0.09 & --1.43 & --1.44 \\
 	 & 	 & 	  & 	Group 3 & & & \\
 47 Tuc  & 0.04  & 13.32 & --0.70 & 0.09 & --0.63 & --0.63 \\
N7006    & 0.10  & 18.00 & --1.48 & 0.04 & --1.40 & --1.41 \\
N3201(a) & 0.25  & 13.61 & --1.56 & 0.10 & --1.48 & --1.49 \\
N6752    & 0.04  & 13.07 & --1.57 & 0.10 & --1.49 & --1.50 \\
N2298    & 0.16  & 15.17 & --1.97 & 0.09 & --1.91 & --1.97 \\
N6397    & 0.24  & 11.62 & --2.02 & 0.07 & --1.96 & --2.02 \\
\hline \hline
\end{tabular}

\hspace{8pt} Model Atmosphere References:\\
MARCS $\equiv$ Gustafsson et al 1975;\\
K-on $\equiv$ Kurucz with overshooting;\\
K-off $\equiv$ Kurucz without overshooting\\
\hspace{8pt} Additional Note:\\
(a) Reddening value is uncertain and variable across face of cluster.
\label{table2}
\end{table}

\section{The Rutledge et al $W'$ versus [Fe/H]$_{\rm II}$ Calibration\label{calibrate}}

Having established revised distance moduli for five key clusters, and
revised metallicities based on Fe~{\sc ii} for sixteen clusters, we 
extended the metallicity scale to other clusters using the reduced EW of 
the IR Ca~{\sc ii} triplet ($W'$) as produced by Rutledge et al (1997a, 
their Table~1).  We refer the reader to KI-03 for further details and
discussion (their \S8, Table~4, and Figures~2 and 4).  The 
coefficients and standard errors ($\sigma$) for the regressions of 
[Fe/H]$_{\rm II}$ on $W'$ derived in KI-03 are as follows:
\begin{enumerate}
\item 
\indent
[Fe/H]$_{\rm II}$ = 0.531$_{(\pm 0.025)}$ $\times$ $W'$ -- 3.279$_{(\pm 0.086)}$   (MARCS)\label{eqn1}
\item 
\indent
[Fe/H]$_{\rm II}$ = 0.537$_{(\pm 0.024)}$ $\times$ $W'$ -- 3.225$_{(\pm 0.082)}$  (Kurucz-on: with conv.\ overshooting)\label{eqn2}
\item 
\indent
[Fe/H]$_{\rm II}$ = 0.562$_{(\pm 0.023)}$ $\times$ $W'$ -- 3.329$_{(\pm 0.078)}$ (Kurucz-off: without conv.\ overshooting)\label{eqn3}
\end{enumerate}

In Table~\ref{table3} we present a compilation of cluster 
[Fe/H]$_{\rm II}$ estimates for which [Fe/H]$_{\rm II} \le$
0.65.  The table includes the metallicity estimates reported in 
Tables~4 and 7 of KI-03, along with estimates for an additional 
39 clusters.   Column (3) lists our best estimate of 
[Fe/H]$_{\rm II}$, based on the new metallicity scale, for a 
given cluster using a MARCS stellar atmosphere model and column 
(4) identifies the method that we employed to derive that 
estimate.  Columns (5 -- 7) show the fit of $W'$ from Rutledge 
et al (or $W'_{est.}$ as estimated below) to the preceding 
equations.

\subsection{$W'$ Extension to Clusters Not Included by Rutledge et al}

For clusters not included in the Rutledge et al compilation (and 
thus not previously reported in KI-03), we estimated a value for
$W'$ on the Rutledge et al system.  We first determined the best 
fit between the Rutledge et al $W'$ measurement to the ZW84 
integrated metallicity index $<$Q39$>$ for clusters in common 
between the studies (where $<$Q39$>$ was not available, we adopted 
the value of Q39 from Zinn 1980).  We then employed this fit to 
interpolate the $<$Q39$>$ (or Q39) value for an additional 37 
clusters to estimate $W'$ on the Rutledge et al system.  This 
value appears in column (8).  We then applied Equations~\ref{eqn1}
--  \ref{eqn3} to derive the estimated [Fe/H]$_{\rm II}$ values 
shown in columns (5 -- 7).  For these clusters, the average error 
in [Fe/H]$_{\rm II}$ is $\sim$0.15~dex, slightly larger than the 
typical error of $\sim$ 0.10~dex quoted for the key clusters with 
values of $W'$ from Rutledge et al.

Thus, the estimates in column (3) of Table~\ref{table3} have been 
obtained by one of the following methods as identified in column 
(4):
\begin{description}
\item[LTG+] metallicity estimates are based on detailed 
	investigation of the results of high resolution
	analyses of individual stars adopting MARCS stellar
	atmosphere models (see KI-03, their \S6, Table~4, 
	and appendix for details).
\item[R-W'] metallicity estimates are based on the extension
	of LTG+ by using $W'$ (the reduced EW of the IR 
	Ca~{\sc ii} triplet from Rutledge et al) and 
	Equation~\ref{eqn1}.
\item[Q39'] metallicity estimates are based on our estimate of
	$W'$ (based on a best fit of the Rutledge et al values
	to $<$Q39$>$ or Q39) and Equation~\ref{eqn1}.
\end{description}
Columns (5 -- 7) display the estimated [Fe/H]$_{\rm II}$ one 
would derive employing the linear correlation of [Fe/H]$_{\rm II}$ 
and $W'$ (or $W'_{est}$) for both MARCS (Equation~1) and Kurucz 
stellar atmosphere models, with and without overshooting 
(Equations~2 and 3, respectively).  Where no LTG+ estimates have 
been made, the results employing Equation~\ref{eqn1} are simply
repeated in column (3).

\begin{table}
\caption{Compilation of Cluster [Fe/H]$_{\rm II}$ Estimates}
\begin{tabular}{ccclcccc}
\hline \hline
{NGC} & {Alt.} & {[Fe/H]$_{\rm II}$} 
                         & {Method} 
                                & {MARCS} & {K-on} & {K-off} & {$W'_{est.}$} \\  
      &  Name &    MARCS &      & (Eq'n 1)& (Eq'n 2)&(Eq'n 3)&  \\    
  (1) &   (2) &      (3) &  (4) &   (5)   &   (6)   &   (7)  & (8) \\    
\hline 
104  & 47 Tuc  & --0.70  & LTG+ & --0.88  & --0.79  & --0.78 & \\ 
288  &         & --1.41  & LTG+ & --1.34  & --1.26  & --1.27 & \\ 
362  &         & --1.34  & LTG+ & --1.27  & --1.20  & --1.21 & \\ 
1261 &         & --1.26  & R-W' & --1.26  & --1.19  & --1.19 & \\ 
     & Eridanus& --1.42  & R-W' & --1.42  & --1.34  & --1.35 & \\ 
1851 &         & --1.19  & R-W' & --1.19  & --1.11  & --1.12 & \\ 
1904 & M79     & --1.64  & R-W' & --1.64  & --1.57  & --1.59 & \\ 
2298 &         & --1.97  & LTG+ & --2.07  & --2.00  & --2.04 & \\ 
2808 &         & --1.29  & R-W' & --1.29  & --1.22  & --1.23 & \\ 
     & Pal 3   & --1.66  & R-W' & --1.66  & --1.59  & --1.61 & \\ 
3201 &         & --1.56  & LTG+ & --1.46  & --1.39  & --1.40 & \\ 
     & Pal 4   & --1.43  & R-W' & --1.43  & --1.35  & --1.37 & \\ 
4147 &         & --1.79  & R-W' & --1.79  & --1.72  & --1.75 & \\ 
4372 &         & --2.29  & R-W' & --2.29  & --2.23  & --2.28 & \\ 
     & Rup 106 & --1.18  & Q39' & --1.18  & --1.10  & --1.10 & 3.960 \\
4590 & M68     & --2.43  & R-W' & --2.43  & --2.37  & --2.43 & \\ 
4833 &         & --2.06  & R-W' & --2.06  & --2.00  & --2.04 & \\ 
5024 & M53     & --2.02  & Q39' & --2.02  & --1.95  & --2.00 & 2.367 \\
5053 &         & --2.41  & R-W' & --2.41  & --2.35  & --2.41 & \\ 
5272 & M3      & --1.50  & LTG+ &$\cdots$ &$\cdots$ &$\cdots$& \\
5286 &         & --1.69  & R-W' & --1.69  & --1.62  & --1.65 & \\ 
5634 &         & --1.92  & Q39' & --1.92  & --1.85  & --1.89 & 2.559 \\
5694 &         & --2.10  & R-W' & --2.10  & --2.04  & --2.08 & \\ 
5824 &         & --1.93  & Q39' & --1.93  & --1.86  & --1.90 & 2.548 \\
5897 &         & --2.09  & R-W' & --2.09  & --2.02  & --2.07 & \\ 
5904 & M5      & --1.26  & LTG+ & --1.32  & --1.24  & --1.25 & \\ 
5927 &         & --0.67  & R-W' & --0.67  & $\cdots$&$\cdots$& \\ 
5946 &         & --1.29  & Q39' & --1.29  & --1.22  & --1.23 & 3.743\\
5986 &         & --1.61  & R-W' & --1.61  & --1.54  & --1.56 & \\ 
     & Pal 14  & --1.61  & R-W' & --1.61  & --1.54  & --1.56 & \\ 
6093 & M80     & --1.76  & R-W' & --1.76  & --1.69  & --1.72 & \\ 
6101 &         & --2.13  & R-W' & --2.13  & --2.07  & --2.12 & \\ 
6121 & M4      & --1.15  & LTG+ & --1.22  & --1.15  & --1.15 & \\ 
6139 &         & --1.61  & Q39' & --1.61  & --1.53  & --1.56 & 3.149\\
6144 &         & --2.10  & R-W' & --2.10  & --2.03  & --2.08 & \\ 
6145 &         & --1.75  & Q39' & --1.75  & --1.68  & --1.71 & 2.885\\
6171 & M107    & --1.10  & R-W' & --1.10  & --1.02  & --1.02 & \\ 
6205 & M13     & --1.60  & LTG+ & --1.58  & --1.51  & --1.53 & \\ 
6218 & M12     & --1.34  & R-W' & --1.34  & --1.25  & --1.26 & \\ 
6229 &         & --1.41  & Q39' & --1.41  & --1.33  & --1.35 & 3.522\\
\hline
     &         &         &   &         &         & Cont'd & ... \\
\hline
\end{tabular}
\label{table3}
\end{table}

\begin{table}
\begin{tabular}{ccclcccc}
\hline \hline
{NGC} & {Alt.} & {[Fe/H]$_{\rm II}$} & {Method} & {MARCS} & {K-on} & {K-off} & {$W'_{est.}$} \\  
      &  Name &    MARCS &      & (Eq'n 1)& (Eq'n 2)&(Eq'n 3)&  \\    
\hline 
6235 &         & --1.39  & R-W' & --1.39  & --1.31  & --1.32 & \\ 
6254 & M10     & --1.51  & LTG+ & --1.48  & --1.41  & --1.43 & \\ 
6266 & M62     & --1.19  & R-W' & --1.19  & --1.11  & --1.12 & \\ 
6273 & M19     & --1.84  & R-W' & --1.84  & --1.77  & --1.79 & \\ 
6284 &         & --1.32  & Q39' & --1.32  & --1.24  & --1.25 & 3.698\\
6287 &         & --2.20  & Q39' & --2.20  & --2.13  & --2.19 & 2.033\\
6293 &         & --2.00  & Q39' & --2.00  & --1.93  & --1.97 & 2.413\\
6304 &         & --0.73  & R-W' & --0.73  & $\cdots$&$\cdots$& \\ 
6316 &         & --0.81  & Q39' & --0.81  & --0.72  & --0.72 & 4.656\\
6325 &         & --1.36  & Q39' & --1.36  & --1.28  & --1.29 & 3.620\\
6333 & M9      & --1.79  & Q39' & --1.79  & --1.72  & --1.75 & 2.811\\
6341 & M92     & --2.38  & LTG+ &$\cdots$ &$\cdots$ &$\cdots$& \\
6342 &         & --0.86  & Q39' & --0.86  & --0.77  & --0.76 & 4.563\\
6352 &         & --0.78  & R-W' & --0.78  & --0.70  & --0.69 & \\ 
6355 &         & --1.33  & Q39' & --1.33  & --1.25  & --1.26 & 3.675\\
6362 &         & --1.15  & R-W' & --1.15  & --1.07  & --1.07 & \\ 
6366 &         & --0.82  & R-W' & --0.82  & --0.74  & --0.73 & \\ 
6388 &         & --0.89  & Q39' & --0.89  & --0.81  & --0.81 & 4.491\\
6397 &         & --2.02  & LTG+ & --2.12  & --2.06  & --2.11 & \\ 
6401 &         & --1.09  & Q39' & --1.09  & --1.01  & --1.01 & 4.119\\
6402 & M14     & --1.41  & Q39' & --1.41  & --1.34  & --1.35 & 3.514\\
6426 &         & --2.43  & Q39' & --2.43  & --2.36  & --2.43 & 1.606\\
6440 &         & --0.72  & Q39' & --0.72  &$\cdots$ &$\cdots$& 4.823\\
6441 &         & --0.85  & Q39' & --0.85  & --0.76  & --0.75 & 4.582\\
6453 &         & --1.46  & Q39' & --1.46  & --1.38  & --1.40 & 3.427\\
6496 &         & --0.78  & R-W' & --0.78  & --0.70  & --0.69 & \\ 
6517 &         & --1.26  & Q39' & --1.26  & --1.18  & --1.19 & 3.808\\
6522 &         & --1.42  & R-W' & --1.42  & --1.35  & --1.36 & \\ 
6535 &         & --1.80  & R-W' & --1.80  & --1.74  & --1.76 & \\ 
6539 &         & --0.87  & Q39' & --0.87  & --0.79  & --0.78 & 4.536\\
6541 &         & --1.83  & R-W' & --1.83  & --1.76  & --1.78 & \\ 
6544 &         & --1.41  & R-W' & --1.41  & --1.34  & --1.35 & \\ 
6553 &         & --0.67  & Q39' & --0.67  &$\cdots$ &$\cdots$& 4.909\\
6558 &         & --1.36  & Q39' & --1.36  & --1.29  & --1.30 & 3.612\\
6569 &         & --0.94  & Q39' & --0.94  & --0.86  & --0.85 & 4.405\\
6584 &         & --1.47  & Q39' & --1.47  & --1.39  & --1.41 & 3.410\\
6624 &         & --0.78  & R-W' & --0.78  & --0.70  & --0.69 & \\ 
6626 & M28     & --1.19  & R-W' & --1.19  & --1.11  & --1.12 & \\ 
6637 & M69     & --0.89  & R-W' & --0.89  & --0.81  & --0.80 & \\ 
6638 &         & --1.04  & R-W' & --1.04  & --0.96  & --0.92 & \\ 
6642 &         & --1.21  & Q39' & --1.21  & --1.13  & --1.14 & 3.890\\
6652 &         & --0.96  & Q39' & --0.96  & --0.88  & --0.87 & 4.372\\
6656 & M22     & --1.71  & Q39' & --1.71  & --1.64  & --1.67 & 2.958\\
\hline
     &         &         &   &         &         & Cont'd & ... \\
\hline
\end{tabular}
\end{table}

\begin{table}
\begin{tabular}{ccclcccc}
\hline \hline
{NGC} & {Alt.} & {[Fe/H]$_{\rm II}$} & {Method} & {MARCS} & {K-on} & {K-off} & {$W'_{est.}$} \\  
      &  Name &    MARCS &      & (Eq'n 1)& (Eq'n 2)&(Eq'n 3)&  \\    
\hline 
     & Pal 8   & --0.81  & Q39' & --0.81  & --0.73  & --0.72 & 4.648\\
6681 & M70     & --1.60  & R-W' & --1.60  & --1.52  & --1.54 & \\ 
6712 &         & --1.10  & R-W' & --1.10  & --1.02  & --1.02 & \\ 
6715 & M54     & --1.47  & R-W' & --1.47  & --1.40  & --1.41 & \\ 
6717 & Pal 9   & --1.27  & R-W' & --1.27  & --1.20  & --1.21 & \\ 
6723 &         & --1.11  & R-W' & --1.11  & --1.03  & --1.03 & \\ 
6752 &         & --1.57  & LTG+ & --1.46  & --1.39  & --1.41 & \\ 
     & Arp 2   & --1.74  & R-W'a& --1.74  & --1.67  & --1.70 & \\
6760 &         & --0.81  & Q39' & --0.81  & --0.73  & --0.71 & 4.654\\
6779 &         & --1.88  & Q39' & --1.88  & --1.81  & --1.85 & 2.637\\
6809 & M55     & --1.85  & R-W' & --1.85  & --1.78  & --1.82 & \\ 
     & Ter 8   & --2.18  & R-W'a& --2.18  & --2.11  & --2.17 & \\
     & Pal 11  & --0.80  & R-W' & --0.80  & $\cdots$&$\cdots$& \\ 
6838 & M71     & --0.81  & LTG+ & --0.82  & --0.74  & --0.73 & \\ 
6864 & M75     & --1.29  & Q39' & --1.29  & --1.22  & --1.23 & 3.743\\
6934 &         & --1.59  & Q39' & --1.59  & --1.52  & --1.54 & 3.178\\
6981 & M72     & --1.42  & R-W' & --1.42  & --1.35  & --1.36 & \\ 
7006 &         & --1.48  & LTG+ & $\cdots$& $\cdots$& $\cdots$ \\ 
7078 & M15     & --2.42  & LTG+ & --2.45  & --2.39  & --2.46 & \\ 
7089 & M2      & --1.56  & R-W' & --1.56  & --1.49  & --1.51 & \\ 
7099 & M30     & --2.33  & R-W' & --2.33  & --2.26  & --2.32 & \\ 
     & Pal 12  & --0.95  & R-W' & --0.95  & --0.87  & --0.86 & \\ 
7492 &         & --1.69  & R-W' & --1.69  & --1.62  & --1.64 & \\ 
\hline \hline
\end{tabular}
\hspace{8pt} Method:\\
\begin{description}
\item[\rm LTG+:] Investigation of individual stellar abundances in key 
	clusters, with new derivations of distance moduli (as shown in 
	Table~\ref{table2}; taken from Table~4 of KI-03).
\item[\rm R-W':] Application of KI-03 Equation~\ref{eqn1} to Rutledge et 
	al determination of $W'$ (taken from Table~7 of KI-03).
\item[\rm R-W'a:] Same as (R-W') but did not appear in Table~7 of
	 KI-03.
\item[\rm Q39':] Application of KI-03 Equation~\ref{eqn1} to $W'_{est.}$.
\end{description}
\end{table}

\section{Some Additional Results and Concerns}

{\it Metal-Rich Clusters ([Fe/H] $>$ --0.65):}  
Most of the globular clusters more metal-rich than those included
in our estimations lie in or near the Galactic bulge, are obscured 
by interstellar dust, and have not been studied extensively at 
high resolution.  Extrapolating the correlations given in 
\S\ref{calibrate} predicts values of [Fe/H]$_{\rm II}$ 0.3 -- 0.5
dex lower than the measured metallicities in the literature.
We refer the reader to further discussion in KI-03 (their \S9).

\hspace{25pt}{\it Determination of [X/Fe] Ratios:}
If Fe is a surrogate for ``metallicity'', then metallicity should 
be represented by the dominant species of Fe.  In the case of giant
stars in globular clusters, the dominant species, by far, is 
Fe~{\sc ii}.  Use of Fe~{\sc i} for this purpose is troublesome due
to possible departures from LTE.   Summarizing the discussion in 
KI-03 (their \S10), we make the following recommendations in the
determinations of [X/Fe] ratios for globular cluster giant stars:
\vspace*{-8pt}
\begin{itemize}
\item C, N, and O are all overwhelmingly in the neutral state.  
Their abundances should be derived from indicators of the abundance
of the neutral species and should be normalized to the abundance of
iron based on Fe~{\sc ii}.
\item Elements such as Ti, Sc, Ba, and Eu, derived from features due
to the singly ionized state, should have their [X/Fe] values 
normalized to the iron abundance based on Fe~{\sc ii}.	
\item Other neutral elements which may suffer from effects of 
over-ionization similar to that of iron, should be normalized to 
abundance of iron based on Fe~{\sc i}.
\end{itemize}

\section{Conclusions}

There exists no ``definitive" set of cluster metallicities that are
systematically reliable on the 0.05~dex level. Any discussion of 
cluster abundances (Galactic or extragalactic) must state clearly the 
underlying assumptions concerning the stellar atmosphere models used;
the adopted T$_{\rm eff}$-scale; what is meant by ``metallicity" 
(eg.\ Fe~{\sc i}? Fe~{\sc ii}? or a mean thereof?); what is the 
method used to derive log~{\it g}; and the origin of 
{\it gf}-values.

\section{Acknowledgements}
This research has made use of NASA's Astrophysics Data System 
Bibliographic Services.  Research by III is currently supported 
by NASA through Hubble Fellowship grant HST-HF-01151.01-A from 
the Space Telescope Science Institute, which is operated by the 
Association of Universities for Research in Astronomy, Inc., 
under NASA contract 5-26555.

We thank the organizers of the Carnegie Observatories Symposium 
on the Origin and Evolution of the Elements for putting 
together such an excellent and stimulating meeting.



\begin{thereferences}{}

\bibitem{}
Alonso, A., Arribas, S., Mart\'inez-Roger, C.\ 1999, \aaps, 140, 261

\bibitem{}
Asplund, M.\ \& Garcia Perez, A.\ 2001, \aap, 372, 601

\bibitem{}
Blackwell, D.\ E., Petford, A.\ D., Arribas, S., Haddock, D.\ J.\ \& Selby, M.\ J.\ 1990, \aap, 232, 396

\bibitem{}
Butler, D.\ 1975, \apj, 200, 68

\bibitem{}
Castelli, F., Gratton, R.\ G., \& Kurucz, R.\ L.\ 1997, \aap, 318, 841 [CG97]

\bibitem{}
Carretta, E., Gratton, R.\ G., Clementini, G.\ \& Fusi Pecci, F.\ 2000, \apj, 533, 215

\bibitem{}
Cohen, J.\ G.\ 1978, \apj, 223, 487

\bibitem{}
Cohen, J.\ G.\ 1979, \apj, 231, 751

\bibitem{}
Cohen, J.\ G.\ 1980, \apj, 241, 981

\bibitem{}
Cohen, J.\ G.\ 1981, \apj, 247, 869

\bibitem{}
ESA 1997

\bibitem{}
Gratton, R.\ G., Carretta, E.\ \& Castelli, F.\ 1996, \aap, 314, 191

\bibitem{}
Gustafsson, B., Bell, R.\ A., Ericksson, K.\ \& Nordlund, A.\ 1975, \aap, 42, 407

\bibitem{}
Houdashelt, M.\ L., Bell, R.\ A.\ \& Sweigart, A.\ V.\ 2000, \aj, 119, 1448

\bibitem{}
Ivans, I.\ I., Kraft, R.\ P., Sneden, C., Smith, G.\ H., Rich, M.\ R.\ \& Shetrone, M.\ 2001, \aj, 328, 1144

\bibitem{}
Kraft, R.\ P.\ \& Ivans, I.\ I.\ 2003, \pasp, 115, 143 [KI-03]

\bibitem{}
Kraft, R.\ P., Sneden, C., Smith, G.\ H., Shetrone, M.\ D., Langer, G.\ E., \& Pilachowski, C.\ A.\ 1997, \aj, 113, 279

\bibitem{}
Kurucz, R.\ L.\ 1992, Rev.\ Mex.\ Astron.\ Astrofis., 23, 181

\bibitem{}
Kurucz, R.\ L.\ 1993, ATLAS9 Stellar Atmosphere Programs and 2kms-1 grid (Kurucz CD-ROM No.\ 13)

\bibitem{}
Nissen, P., Primas, F., Asplund, M. \& Lambert, D.\ 2002, \aap, 390, 235

\bibitem{}
Pilachowski, C.\ A., Sneden, C., Kraft, R.\ P.\ \& Langer, G.\ E.\ 1996, \aj, 112, 545

\bibitem{}
Reid, I.\ N.\ 1997, \aj, 114, 161

\bibitem{}
Rutledge, G., Hesser, J.\ \& Stetson, P.\ 1997a, \pasp, 109, 907

\bibitem{}
Rutledge, G., {\it et al.}\ 1997b, \pasp, 109, 883

\bibitem{}
Sneden, C., Kraft, R.\ P., Prosser, C.\ \& Langer, G.\ E.\ 1992, \aj, 104, 2121

\bibitem{}
Th\'evenin, F.\ \& Idiart, T.\ P.\ 1999, \apj, 521, 753 [TI99]

\bibitem{}
Zinn, R.\ 1980, \apjs, 42, 19

\bibitem{}
Zinn, R.\ \& West, M.\ J.\ 1984, \apjs, 55, 45 [ZW84]

%
%
%

\end{thereferences}
\end{document}